\documentstyle[12pt,epsfig]{article}
\begin{document}
\renewcommand{\arraystretch}{1.5}


\pagestyle{empty}
\rightline{UG-4/96}
\rightline{HUB-EP-96/10}
\rightline{hep-th/9604168}
\rightline{April 1996}
\rightline{revised version}
\vspace{1truecm}
\centerline{\bf  Intersecting $D$--Branes in Ten and Six Dimensions}
\vspace{1truecm}
\centerline{\bf Klaus Behrndt}
\vspace{.4truecm}
\centerline{{\it Institut f\"ur Physik, Humboldt-Universit\"at}}
\centerline{{\it10115 Berlin}}
\centerline{{\it Germany}}
\vspace{1truecm}
\centerline{\bf Eric Bergshoeff and Bert Janssen}
\vspace{.4truecm}
\centerline{{\it Institute for Theoretical Physics}}
\centerline{{\it Nijenborgh 4, 9747 AG Groningen}}
\centerline{{\it The Netherlands}}
\vspace{2truecm}
\centerline{ABSTRACT}
\vspace{.5truecm}
We show how, via $T$--duality, intersecting $D$-Brane configurations
in ten (six) dimensions can be obtained from the elementary $D$-Brane
configurations by embedding a Type IIB $D$-Brane into a Type IIB
Nine--Brane (Five--Brane) and give a classification of such configurations.
We show that only a very specific subclass of these configurations
can be realized as (supersymmetric) solutions to the equations of motion
of IIA/IIB supergravity.

Whereas the elementary $D$--brane solutions in $d=10$ are characterized
by a single harmonic function, those in $d=6$ contain two independent
harmonic functions and may be viewed as the intersection of two $d=10$
elementary $D$-branes. Using string/string/string triality in 
six dimensions we show that the heterotic version of the elementary
$d=6$ $D$-Brane solutions correspond in ten dimensions to intersecting 
Neveu-Schwarz/Neveu-Schwarz (NS/NS) strings or five-branes
and their $T$--duals. We comment on the implications of our results
in other than ten and six dimensions.

\vfill\eject
\pagestyle{plain}
 
\section{Introduction}
 
Recent developments have shown that an important role in
understanding nonperturbative string theory 
is played by $p$--Brane solutions of the string effective action whose
charge is carried by a single Ramond/Ramond (RR) gauge field. 
These solutions correspond to open string states whose endpoints are
constrained to live on a $p+1$--dimensional worldvolume and are
known as Dirichlet $p$--Branes or simply $D$--Branes \cite{Pol1}
(for a review, see \cite{Pol2}). More precisely, a Dirichlet $p$--Brane
in ten dimensions is an open string state which satisfies Dirichlet
boundary conditions for the $9-p$ transverse directions and Neumann
boundary conditions for the $p+1$ worldvolume directions. Since under
$T$--duality\footnote{For a review, see \cite{Gi1}.}
Dirichlet and Neumann boundary conditions are
interchanged it follows that all Dirichlet $p$--Branes 
$(p=0,\cdots,9)$ are $T$--dual versions of each other \cite{Pol2}.
It is natural that this $T$--duality is also realised on
the $p$--Brane solutions and indeed this has shown to be the case
for all values of $p$ with $0\le p\le 9$ \cite{Be1}. 

The elementary Dirichlet $p$--Brane solutions in ten dimensions are
characterized by a single function $H_p$ that depends on the $9-p$
transverse coordinates and is harmonic with respect to these variables.
The metric for all values of $p\ (0\le p\le 9)$ is given by
\begin{equation}
\label{eD10}
ds^2 = H_p^{-1/2}ds^2_{p+1} +
H_p^{1/2}ds^2_{9-p}\, .
\end{equation}
For even (odd) $p$ this metric corresponds, together with
certain expressions for the dilaton and the RR gauge field, 
to a solution of IIA (IIB) supergravity. The only
nontrivial $T$--duality rule involving the metric is
given by
\begin{equation}
{\tilde g}_{xx} = 1/ g_{xx}\, .
\end{equation}
where $x$ labels the isometry direction over which the duality is
performed. Clearly, under this duality transformation the metric of
a Dirichlet $p$--Brane becomes that of a $p+1$--Brane if the
duality is performed over one of the transverse directions of
the $p$--Brane. In other words, one of the transverse directions
of the $p$--Brane has become a worldvolume direction of the $p+1$--Brane.
We assume here that the harmonic function $H_p$
is independent of the particular transverse direction which is dualized
or, alternatively, that we consider a periodic array of $p$--Brane solutions.
We therefore may write 

\begin{equation}
{\tilde H}_p = H_{p+1}\, .
\end{equation}
Conversely, a $p+1$--Brane becomes a $p$--Brane if the duality
is done over one of the worldvolume directions of the $p+1$--Brane.
In this case, in order to establish a duality between the two
solutions, we assume that after duality the harmonic
function $H_{p+1}$ becomes dependent on this particular worldvolume direction.
It is in this sense that we may write 

\begin{equation}
{\tilde H}_{p+1} = H_p\, .
\end{equation}
Strictly speaking the $T$ duality rules can only
be applied as solution generating transformations to
construct $p+1$--Brane solutions out of $p$--Brane solutions
and not the other way around. Therefore the 0--Brane leads, 
via $T$--duality, to all other $D$--Brane solutions. 
{\it A priori} it is not guaranteed that one can apply $T$--duality also
as a solution-generating transformation
to construct $p$--Brane solutions out of $p+1$--Brane solutions.
In the case of the $D$--brane solutions the $T$--duality
does generate new dual solutions but, as we will see later, this
is not true for other configurations. For the general case one
must check by hand whether the dual metric indeed is a new solution. 
In this paper we will use $T$--duality only as a tool to construct a
natural Ansatz for a class of solutions.

A particularly
interesting case, that will play an important role later,
is the nine--Brane solution which has no transverse
directions and whose worldvolume
is ten--dimensional Minkowski spacetime. Therefore, all Dirichlet 
$p$--Branes are, via $T$-duality in the $9-p$ transverse directions,
$T$--dual to flat spacetime. Conversely, out of flat spacetime
we can construct all other $D$--Brane solutions via $T$--duality.
This can be done in the following way. First
we write the nine--Brane solution as

\begin{equation}
ds^2 = H_9^{-1/2}ds^2_{10}\, ,
\end{equation}
where $H_9$ is a constant (which is related to the
spacetime volume). 
To obtain the 8-Brane solution, we 
dualize in one of the worldvolume directions, say $x^9$, and
assume that after duality $H_9$ becomes dependent on $x^9$,
i.e.~${\tilde H}_9 = H_8$. One thus obtains the 8--Brane solution.
Similarly, 
one can obtain all other $D$--Brane solutions. We conclude that not
only does the 0--Brane, via $T$--duality, leads to all other Dirichlet
$p$--Branes with $0 < p \le 9$ but also does the nine--Brane lead,
via $T$--duality, to all remaining Dirichlet $p$--Branes with $0\le p < 9$.

It is natural to consider bound states 
of (orthogonal) intersections of $D$--Branes. 
Such multiple $D$--Brane configurations
have been considered in \cite{Wi1,Pol2}\footnote{
Generalizations of $D$--Branes
where an open $p$--Brane ends on a $q$--Brane have been
discussed in \cite{St1,To1}.}. It turns out that, whereas the 
elementary $D$--Brane solutions are described by a single harmonic
function, the solution corresponding to $q$ intersecting $D$--Branes 
contains $q$ independent harmonic functions. Thus, these solutions may
be considered as compositions of the elementary solutions. For
simplicity, we will first restrict ourselves to solutions with
two independent harmonic functions. Several examples of intersecting
solutions in ten dimensions (not necessarily $D$--Branes but also 
NS/NS solutions)
with more than one harmonic function have been given in the
literature. The configuration given in \cite{Kh1} describes two NS/NS 5-Branes
intersecting over a string. The solution given in eq.~(2.5) of \cite{Duff}
describes 3 NS/NS 5--Branes intersecting over a string, while the solution
given in eq.~(5.2) of the same reference describes the intersection of
a fundamental string with two NS/NS 5--Branes. 
The chiral null model of \cite{Ho1} describes 
the composition of a NS/NS string and its $T$--dual while the
the solution given in \cite{Cv1,Ts1} 
describes a fundamental string and its T--dual lying within a NS/NS
5-Brane and its T--dual. Finally, the solution of \cite{Ca1} describes
a Dirichlet 1--Brane inside a Dirichlet 5--Brane. 

Recently, intersecting $p$--Brane solutions have also been considered
in eleven dimensions in which case they are referred to as
intersecting $M$--Branes. More explicitly, it has been shown in 
\cite{Pa1} that the solutions of \cite{Gu1} that break more than
one half of the eleven--dimensional supersymmetry can be considered
as intersections of the elementary 2--Branes \cite{Du1} and 
5--Branes \cite{Gu1}. This analysis has been extended in \cite{Ts2}
where intersections with independent harmonic functions for each
elementary constituent has been considered. The discussion of
\cite{Ts2} was also applied to ten dimensions where several
examples of solutions describing intersecting $D$--Branes were given.
For all these solutions,
the general rule seems to be that the intersection of $q$ solutions
preserves at most $(1/2)^q$ supersymmetry.

It is the purpose of this work to give a systematic and unified
treatment of solutions describing intersecting Dirichlet $p$--Branes
in ten and six dimensions. For this purpose we first consider the
elementary $D$--Brane solutions in general dimensions\footnote{ Since
the zero-modes of $D$--Branes are described by a worldvolume vector
multiplet we should restrict ourselves to those dimensions in which
there is a Bose--Fermi matching on the worldvolume using vector
multiplets. This restricts ourselves to $d=3,4,6,10$ dimensions
\cite{Du3}. Note that only in $d=6,10$ we can distinguish between IIA
and IIB theories.}. We define a solution to be an elementary Dirichlet
$p$--Brane if (i) it is charged with respect to one or more RR gauge
fields, (ii) the metric contains a single function $H_p$ which is
harmonic with respect to the transverse variables and has the property
that the worldvolume and transverse directions are multiplied with
{\it opposite} powers of this function and (iii) the solution has
vanishing scalar except for a possible dilaton. From the requirement
(ii) it follows that the metric of all elementary $D$-Branes are
related via $T$--duality, a property which we expect from an open
string state with mixed Dirichlet/Neumann boundary conditions. The
requirement (iii) ensures that the solutions are connected to flat
space time with constant scalars.

To determine the metric of an elementary $D$--Brane solution it is
convenient to first consider those $D$-Branes that have vanishing dilaton.
The other $D$--Brane solutions are then obtained via $T$--duality.  It
is not too difficult to see, using the results of \cite{Lu1}, that
such solutions only exist for $p= (D-4)/2$ and that the metric of such
solutions is given by
\begin{equation}
ds^2 = H_p^{-4/(D-2)}ds^2_{p+1} + 
H_p^{4/(D-2)}ds^2_{D-p-1}\, .
\end{equation}
For $d=10$ this reproduces the metric of (\ref{eD10}). Instead, we
see that for $d=6$ the elementary $D$--Brane solutions are given by
\begin{equation}
\label{eD6}
ds^2 = H_p^{-1}ds^2_{p+1} + 
H_p ds^2_{5-p}\, .
\end{equation}
We conclude that the elementary $D$--Branes in six dimensions cannot
be obtained from a dimensional reduction of elementary $D$--Branes in
ten dimensions since the two solutions have different powers of 
harmonic functions in their respective metrics\footnote{The
reduction of the elementary $d=10$ $D$--Branes to six dimensions
lead to solutions with extra nonvanishing scalars since the metric
in the compactified directions is nontrivial.}.
In this paper we will show, following a suggestion of \cite{Pa1,Ts2},
how all the elementary $D$--Brane
solutions in six dimensions have a higher--dimensional interpretation 
in terms of intersecting $D$--Branes in ten dimensions.
Furthermore we will construct, by using string/string/string triality 
\cite{Du2,Be2},
the heterotic analogue of the six--dimensional $D$--Branes and
show that their ten--dimensional origin is given by 
ten--dimensional intersecting 
Neveu-Schwarz/Neveu-Schwarz (NS/NS) strings or five-Branes
and their $T$--duals.

The organization of this paper is as follows. In section 2 we will
first classify all possible intersecting configurations consisting of two 
elementary $D$--Branes in ten dimensions and briefly discuss the
extension to more than two intersecting $D$--Branes.
After that we will show that only a specific subclass of these
configurations can be (supersymmetric) solutions to the
equations of motion of IIA/IIB supergravity.
In section 3 we will use these results to explain the ten--dimensional
origin of the $D$--Brane solutions in six dimensions. 
Furthermore we will
construct the heterotic analogue of the $d=6$ $D$--Branes and
explain their ten--dimensional origin.
Finally, in the last section
we comment on the implications of our results
in other then ten and six dimensions.

\section{$D$-Branes and their intersections in $d=10$}

In this section we discuss the intersection of two $D$--Branes in
ten dimensions. 
To construct configurations which describe intersections, a 
crucial role is played by $T$--duality.
Let us first explain, via an example, the main idea of this construction.
After that we will give a more systematic treatment.
Consider an elementary Dirichlet $p$--Brane with
$1 \le p \le 7$ odd and metric given in (\ref{eD10}). We embed
this $D$--Brane in flat spacetime, i.e. a nine--Brane,
and write
\begin{equation}
ds^2 = {1\over \sqrt {H_pH_9}}ds^2_{p+1} - \sqrt {{H_p\over H_9}}
ds^2_{9-p}\, ,
\end{equation}
where $H_9$ is the (constant) harmonic of the nine--Brane.  In the
table at the end we have indicated this embedding by the $\|$ symbol,
e.g.~we write $1\| 9, 3\| 9$, etc. We next may perform a $T$--duality
transformation in two different ways.  The first possibility is that
we perform the $T$--duality over one of the transverse directions of
the $p$--Brane, say $x^9$, which from the nine--Brane point of view is
a worldvolume direction.  In this way we obtain the
configuration\footnote{The expressions for the gauge fields and dilaton can
easily be obtained by applying the Type II $T$--duality rules of
\cite{Be3}. They will be given below for the general case.}
\begin{equation}
\label{new1}
ds^2 = {1\over \sqrt {H_{p+1}H_8}} ds^2_{p+1} - \sqrt {{H_{p+1}\over 
H_8}}ds^2_{8-p} - \sqrt {H_8 \over H_{p+1}} (dx^9)^2\, ,
\end{equation}
where we have used that $\tilde {H}_p = H_{p+1}$ and $\tilde {H}_9
= H_8$. We thus end up with a configuration describing the intersection of
a $(p+1)$--Brane with an 8--Brane via a $p$--Brane. 
In the table these new configurations
are indicated one step to the right with respect to the
original configuration, e.g. 
\begin{equation}
1\| 9 \rightarrow 2\bot 8\, .
\end{equation}
The second possibility is that we perform the $T$--duality over one
of the worldvolume directions of the $p$--Brane, say $x^1$. 
This leads to the configuration
\begin{equation}
\label{new2}
ds^2 = {1\over \sqrt {H_{p-1}H_8}} ds^2_{p} - \sqrt {{H_{p-1}\over 
H_8}}ds^2_{9-p} - \sqrt {H_{p-1}H_8} (dx^1)^2\, ,
\end{equation}
where we have used that $\tilde {H}_p = H_{p-1}$ and $\tilde {H}_9
= H_8$. This configuration describes a $(p-1)$--Brane 
embedded into an 8--Brane. 
In the table these new configurations are indicated one step above the
original configuration, e.g.
\begin{equation}
1\| 9 \rightarrow 0\| 8 \, .
\end{equation}
Clearly, from the configurations (\ref{new1}) and (\ref{new2}) one can
obtain further intersecting configurations by performing $T$--duality over
different directions.  This leads us to consider the following
class of metrics with two independent harmonic functions $H_{p+r},
H_{p+s}$ describing the intersection of a $(p+r)$--Brane with a
$(p+s)$--Brane via a $p$--Brane:
\begin{equation}
\label{general}
ds^2 = {1\over \sqrt {H_{p+r}H_{p+s}}}ds^2_{p+1} - \sqrt {{H_{p+r}\over 
H_{p+s}}}
ds^2_s - \sqrt {{H_{p+s}\over H_{p+r}}}
ds^2_r - \sqrt {H_{p+r}H_{p+s}} ds^2_{9-p-r-s} \ .
\end{equation}
The dilaton is given by
\begin{equation}
\label{dilaton}
e^{-2 \phi} = (H_{p+r})^{\frac{p+r-3}{2}}  (H_{p+s})^{\frac{p+s-3}{2}} \ .
\end{equation}

Using the terminology of \cite{Pa1,Pa2} we call the first $p+1$
coordinates the worldvolume coordinates of the intersection.
The next $s$ and $r$ directions are
the ``relative transverse'' directions with respect to the
$p+r$--Brane and $p+s$--Brane, respectively. Finally, the last 
$9-p-r-s$ are the
``overall transverse'' directions. 
We recover the metric of the elementary Dirichlet $(p+r)$ and $(p+s)$-Brane
by putting $H_{p+s}$ and $H_{p+r}$ equal to one, respectively. 
We will denote the configuration given in (\ref{general})
as $(p+r) \times (p+s)$.
Note that if $r$ or $s$ is equal 
to zero the configuration describes the embedding of one $D$-Brane into another
one. We will indicate this case as $p\parallel(p+s)$. 
On the other hand, if both $r$ and $s$ are different from
zero, the configuration (\ref{general}) describes two $D$-Branes which are 
orthogonally intersecting over a 
$p$-Brane. We will indicate this situation as $(p+r)\perp(p+s)$.

The labels 
$p,r$ and $s$ in the general configuration (\ref{general})
have to fulfill certain conditions: first of all $p+r+s\leq 9$
for the obvious reason that we only have 9 spatial dimensions to fill. 
Further we only want to combine objects which come from the same theory (IIA
or IIB), so $r+s$ has to be an even number $2n$, where $n$ labels 
different classes of configurations, as indicated in the table.

In general the configuration $(p+r)\times (p+s)$ transforms under $T$--duality
as follows. One possibility is that
\begin{equation}
\label{case1}
(p+r)\times(p+s) \rightarrow (p+(r \pm 1)) \times (p+(s \mp 1))\, ,
\end{equation}
if under the duality 
a relative transverse direction of one object becomes a relative transverse
direction of the other object. In this case we move horizontally
in the table. The second possibility is that we have
\begin{equation}
\label{case2}
(p+r)\times(p+s) \rightarrow ((p\pm 1)+r)\times((p\pm 1) +s)\, ,
\end{equation}
if under duality 
an overall transverse direction becomes an intersecting one or vice versa.

In either case (\ref{case1}) or (\ref{case2}) the number $r+s=2n$
remains constant, so that $n$ can be used to label the four different
different classes given in the table.  Within each class we may move
via $T$-duality in the way descriped above.  To go from one class to
another we first rewrite the $(p\parallel 9)$ element of the class as
an elementary $D$--Brane $p$.  Next we transform $p$ into $q$ under
$T$--duality and write $q$ as $q\|9$ which is the element of a
different class.  There is one subtlety here. It only makes sense to
embed a $p$--Brane into a 9--Brane for $(p = {\rm odd})$-Branes. The
reason is that one should view the nine--Brane as a IIB solution since
under $T$--duality it is transformed into a IIA solution (the
8--Brane). This implies that it can only intersect with another IIB
solution.

In conclusion, we can summarize all possible intersections of two
$D$-Branes in 10 dimensions as given in the table: the blocs at the
bottom are the elementary $D$-Branes, and on each $(p = {\rm
odd})$-solutions we can build a tower of intersecting $D$-Brane
configurations, labeled by the integer $n$. Within each tower we can move
from one configuration to the horizontally and vertically neighbouring ones
via the different possible $T$-dualities: horizontal moves, in which
$p$ remains constant correspond to dualizing relative transverse
coordinates, while vertical moves keep $r$ and $s$ constant and
correspond to dualizing worldvolume directions into overall transverse
ones or vice versa.

The explicit expressions for the RR fields follow from the T-duality
(for  more details see \cite{Be1}). Alternatively, their expression can easily
be obtained by the requirement that, if one of the harmonic functions is
set equal to one, the intersecting configuration should reduce to one
of the $D$--brane solutions discussed in the introduction.
The explicit form of the RR gauge fields is most easily given by 
using a formulation where the magnetic configurations are descibed
by magnetic (dual) potentials. This leads us to consider the Lagrangian

\begin{equation}
\label{L}
{\cal L} = \sqrt {-g}\left\{e^{-2\phi}\biggl [ R - 4(\partial\phi)^2\biggr ]
+ {(-)^{p+r+1}\over 2(p+r+2)!}F^2_{(p+r+2)}
+ {(-)^{p+s+1}\over 2(p+s+2)!}F^2_{(p+s+2)}\right\}\, ,
\end{equation}
where it is understood that in the field equations one imposes
the constraint that $F_{(8-p)}$ is the dual of $F_{(p+2)}$. In particular,
$F_{(5)}$ is selfdual. Pseudo-Lagrangians of this form have been discussed in
\cite{BBO}. It is also understood that the two kinetic terms for the
gauge fields become identical if $r=s$.

We next distinguish three different cases:
\bigskip

\noindent {\bf Case (1)}\ \ Both harmonic functions depend on the overall
transverse directions. The RR gauge fields are given by

\begin{equation}
\label{F_1}
F^{(1)}_{0\cdots p1\cdots ri} = \partial_i H_{p+r}^{-1}\, ,
\hskip 1.5truecm
F^{(2)}_{0\cdots p1\cdots si} = \partial_i H_{p+s}^{-1}\, .
\end{equation}
\bigskip

\noindent {\bf Case (2)}\ \ The function
$H_{p+r}$ depends on the overall
transverse directions whereas $H_{p+s}$ depends on the relative transverse
directions. The RR gauge fields are given by

\begin{equation}
\label{F_2}
F^{(1)}_{0\cdots p1\cdots ri} = H_{p+s}^{\alpha}\partial_i H_{p+r}^{-1}\, ,
\hskip 1.5truecm
F^{(2)}_{0\cdots p1\cdots sr} = \partial_r H_{p+s}^{-1}\, .
\end{equation}
\bigskip

\noindent {\bf Case (3)}\ \ Both harmonic functions depend on the relative
transverse directions. The RR gauge fields are given by

\begin{equation}
\label{F_3}
F^{(1)}_{0\cdots p1\cdots rs} = H_{p+s}^{\alpha}\partial_s H_{p+r}^{-1}\, ,\hskip 1.5truecm
F^{(2)}_{0\cdots p1\cdots sr} = H_{p+r}^{\beta}\partial_r H_{p+s}^{-1}\, .
\end{equation}
\bigskip

\noindent
As we will see below, one can also consider more general cases
which are compositions of these three different cases.
For simplicity we will often, if not necessary, omit the expressions 
for the
RR   fields\footnote{Note that the RR gauge fields are contributing
to the Wess-Zumino part of the world volume actions. As for the 
fundamental string the world volume actions provide the source
terms of the classical solution.}.
Note that the $\alpha$ in case (2) and the $\alpha, \beta$ in case (3)
are arbitrary (real) parameters that cannot be fixed by the Bianchi
identities. They can be determined by the $T$-duality.
Alternatively, we will determine them below by the equations of motion.
 
Sofar, we have only applied $T$-duality to generate the general
Ansatz (\ref{general}), (\ref{dilaton}), (\ref{F_1}-\ref{F_3})
for intersecting $D$--brane configurations. 
Our next task is to determine which of these configurations
corresponds to a (supersymmetric) solution of the Lagrangian (\ref{L}).
Substituting our Ansatz into the vector 
field and dilaton equation\footnote{The case that only
intersecting 3--Branes are involved is special since for this
case the dilaton equation is trivially satisfied. By applying $T$--duality
one can relate this case to the other cases and show that the same
restrictions as given below apply.}, we see that {\it case $(1)$ 
can only be a solution for
$n=2$, case $(2)$ for $n=2$ and $\alpha = 0$ while case $(3)$ 
requires that $n=4$ and $\alpha=\beta=1$.} 
Furthermore, it turns out that the cases (1) and (2) can naturally 
be combined into a more general configuration
where $H_{p+r}$ only depends on the overall transverse
directions, as before, but where $H_{p+s}$ is given by the sum
of two harmonics $H^{(1)}_{p+s}, H^{(2)}_{p+s}$, which depend on the overall
and relative transverse directions, respectively, i.e.

\begin{equation}
H_{p+s}(x^i,x^r) = H^{(1)}_{p+s}(x^i) + H^{(2)}_{p+s}(x^r)\, .
\end{equation}

We will now investigate the supersymmetry of these solutions.
It has been shown in \cite{Pol2} that only those interesections
can be supersymmetric (1/4 of the supersymmetry is unbroken)
that satisfy the condition that $r+s = 0\ mod\ 4$, i.e. $n=2$ or $4$.
In our language
this goes as follows. For a single $D$-brane the supersymmetry condition
follows from $\delta\lambda = 0$, where $\lambda$ is the
spinor in the IIA/IIB supergravity multiplet,  and $\delta\lambda$ 
(in the string frame) is given by

\begin{equation}
\delta\lambda = 
\gamma^\mu (\partial_\mu\phi)\epsilon + {1\over 4}{3-p\over
(p+2)!}e^\phi
F_{\mu_1\cdots \mu_{p+2}}\gamma^{\mu_1\cdots \mu_{p+2}}\ 
\epsilon_{(p)}^\prime = 0\, ,
\end{equation}
where $\epsilon_{(p)}^\prime = \epsilon$ for $p=0,4,8$;
$\epsilon_{(p)}^\prime = \gamma_{11}\epsilon$ for $p=2,6$;
$\epsilon_{(p)}^\prime = i\epsilon$ for $p=-1,7$ and
$\epsilon_{(p)}^\prime =
i\epsilon^\star$ for $p=1,5$\footnote{For $p=3$ the supersymmetry
condition does not
follow from $\delta\lambda = 0$ (there is no dilaton and the 4--form gauge
field is absent in $\delta\lambda$). For that case one
has to consider the supersymmetry rule of the gravitino.}.
Substituting the $D$--Brane solution into the above equation leads
to the condition

\begin{equation}
\epsilon + \gamma_{01\cdots p}\ \epsilon_{(p)}^\prime = 0\, , 
\end{equation}
i.e.~half of the supersymmetry is broken.

Now consider the intersection of a $(p+r)$-Brane with a $(p+s)$-Brane.
Then the two supersymmetry conditions corresponding to the $(p+r)$-Brane
and $(p+s)$-Brane are given by

\begin{eqnarray}
\epsilon + \gamma_{01\cdots p+r}\epsilon^\prime_{(p+r)} = 0\, ,\nonumber\\
\epsilon + \gamma_{01\cdots p+s}\epsilon^\prime_{(p+s)} = 0\, ,
\end{eqnarray}
respectively. Combining the two supersymmetry conditions we get

\begin{equation}
\epsilon_{(p+r)}^\prime = (-)^{{1\over 2}r(r+1)} \gamma_{r+s}
\epsilon^\prime_{(p+s)}\, .
\end{equation}
We now distinguish four cases in which the two spinors in the above
equation are given by $(\epsilon,\epsilon), (\epsilon,\gamma_{11}\epsilon),
(i\epsilon,i\epsilon)$ or $(i\epsilon, i\epsilon^\star)$, respectively.
All four cases lead to the consistency condition that $\gamma_{r+s}^2 = 1$,
or

\begin{equation}
n = 2\ \ \ {\rm or}\ \ \ 4\, ,
\end{equation}
thereby reproducing the condition of \cite{Pol2}.

We next extend this analysis and consider the
Killing spinor equation that follows from
$\delta\lambda = 0$ for the case that we substitute the
complete intersecting configuration and not only the separate $D$--Brane
configurations. In the string--frame we obtain the following 
equation from $\delta\lambda = 0$:

\begin{eqnarray}
\gamma^\mu (\partial_\mu\phi)\epsilon &+&{1\over 4}(3-p-r)e^\phi
F^{(1)}_{0\cdots p+r,\mu}\gamma^{0\cdots p+r,\mu}\ 
\epsilon_{(p+r)}^\prime\, +\nonumber\\
&+&{1\over 4}(3-p-s)e^\phi
F^{(2)}_{0\cdots p+s,\mu}\gamma^{0\cdots p+s,\mu}\ 
\epsilon_{(p+s)}^\prime = 0\, .
\end{eqnarray}
Substituting the explicit form of the 
general intersecting configuration (\ref{general}), (\ref{dilaton}),
(\ref{F_1}-\ref{F_3})
into the above Killing spinor equation leads, for case (1) to $n=2$, for
case (2) to $n=2, \alpha= 0$ and for case (3) to 
$n=4, \alpha=\beta= 1$. 
This nicely agrees with our earlier finding that only
these configurations can be solutions to the equations of motion.
It is not clear to us what the role is of the intersecting $D$-Branes
with $n=1,3,$ and whether they can be represented as some kind of
solution.

Finally, to construct solutions with more than two $D$--Branes, one
can follow the same procedure.  We take the  configuration
(\ref{general}) for two intersecting $D$--Branes with $n=2,4$ and
for the case that
both Branes belong to the IIB theory and embed this solution into a
Nine--Brane. Next, one should apply $T$--duality in the different
possible directions to obtain the class of configurations describing three
intersecting $D$--Branes. This process can be repeated. It would be
interesting to determine which of these configurations can be
(supersymmetric) solutions to the equations of motion.

\section{Elementary $D$-Branes in 6 dimensions}

\noindent
In this section we describe the compactification of 2 intersecting
$D$-Branes. In 6 dimensions we are interested in a basic set of
non-intersecting $D$-Branes, which means that we have to compactify
over the relative transversal space.  Since the internal space has 4
dimensions we have to consider the class $n=2$.  From the table we
conclude that there are 3 families given by the 3 columns in the $n=2$
class, e.g. for a common 0-Brane ($p=0$), we have $0 \| 4$, $1 \bot 3$
and $2 \bot 2$. Since we are compactifing over a $K3$ we
have to discard the second family. There is no place for a string
inside the $K3$ (no 1-cycles). Thus we are left with the other two
intersections.  Next, we have to discuss how to put these objects into
the $K3$.  The $0 \| 4$ object is clear, the 0-Brane is unchanged and
the 4-Brane wraps around the whole $K3$. As a result we get one object
defined by two charges. The two membranes intersecting over a point
($2 \bot 2$) have two 2-cycles that have no common point. Obviously
the two membranes have to wrap around two of the 22 2-cycles. Naively
one would think that this gives us 22 possibilities where every
possibility is defined by two charges. But instead, we have to take
into account that the 2-cycles in the $K3$ pick up only the self- or
antiself-dual part of the 2 membranes. Hence every possibility is
related only to one charge. In total, the 6d solution
is defined by 24 charges, that have to form a $O(4,20)$ vector.

Therefore, a natural ansatz for the Type IIA $D$-0-Brane is given by
\begin{equation} \label{6dIIA}
ds^2_{IIA}=  e^{2U} dt^2 - e^{-2U} ds^2_5 \ \  , \ \  
e^{-4U} = e^{4\phi} =(|\vec{\chi}^R|^2 - |\vec{\chi}^L|^2) \ ,
\end{equation}
where $\vec{\chi}^{R/L}$ is a harmonic vector in the $O(4,20)$
space. To clarify this formula we consider the case $0\| 4$,
where we have only two harmonic functions
\begin{equation}
\left(\begin{array}{c} \chi^R \\ \chi^L \end{array} \right)
 = \frac{1}{\sqrt{2}} \left( 
 \begin{array}{c} H_0  + H_4   \\ 
 H_0 - H_4  \end{array} \right) \ ,
\end{equation}
where $H_0$ and $H_4$ are related to the two intersecting $D$-Branes
in 10 dimensions with the metric given by 
\begin{equation}
ds^2 = \frac{1}{\sqrt{H_0 H_4}} dt^2 - \sqrt{H_0 H_4} ds_5^2 - 
\sqrt{\frac{H_0}{H_4}} ds_4^2 \ .
\end{equation}
Wrapping the 4 relative transversal space around the $K3$ yields
the following 6-dimensional metric
\begin{equation}
ds^2 = \frac{1}{\sqrt{H_0 H_4}} dt^2 - \sqrt{H_0 H_4} ds_5^2 \ ,
\end{equation}
which is a special case of (\ref{6dIIA}) and for $H_0=H_4$ it
coincides with (\ref{eD6}).

As in 10 dimensions the electric gauge fields are proportional to the
inverse power of a harmonic function and an $O(4,20)$ covariant ansatz
is (including the 16 left moving modes)
\begin{equation}
\vec{A}_0 = e^{4U} 
\left(\begin{array}{c} \vec{\chi}^R \\ \vec{\chi}^L \end{array} \right) \ .
\end{equation}
Also, there are scalar fields which are given by the matrix ${\cal M}$
that parameterizes the $O(4,20)$ space. To find this matrix
and to proof that this is really a solution of the low energy effective
action we have to compare the $D$-Brane solution with known solutions on
the  heterotic side (see below).

To find the dyonic $D$-1-Brane we have to compactify the intersections
$1\|5$, $3\bot 3$ of the class $n=2$. Again the internal structure of
the $K3$ yields a result in 6 dimensions which has an $O(4,20)$
structure
\begin{equation} 
\begin{array}{ccc}
ds^2_{IIB}=  e^{2U} ds^2_2
- e^{-2 U}ds^2_4  \quad , \quad e^{2\phi} = 1 \quad , \quad
\vec{B}_{0 1} = \vec{A}_{0} \ ,
\end{array}
\end{equation}
where the other components are defined via the (anti)self-duality
condition: $\vec{H} = ({\cal M} \cdot {\cal L} ^* H)$ with the matrix
${\cal L}$ defining the metric in the $O(4,20)$ space (${\cal M}$ 
is the same form heterotic, IIA and IIB). 

The magnetic 2--Branes can be obtained by reducing the $2 \| 6$ and
$4\bot 4$ solutions. The $D$-2-Brane in six dimensions has the form 
\begin{equation} 
\begin{array}{c}
ds^2_{IIA}=  e^{2U} ds^2_3
- e^{-2U} ds^2_3 \quad , \quad e^{2\phi} = e^{2U} \quad , \quad
\vec{F}_{ij} = \sqrt{2} \epsilon_{ijm} \partial_m \vec{\chi} \ .
\end{array}
\end{equation}
The higher Branes are not asymptotically flat.  For instance the
Dirichlet 3--Branes in six dimensions are obtained from the reduction
of the $3\| 7$ and $5 \bot 5$ solutions.  The 4-Brane is related to a
cosmological constant and, finally, the 6d space time can be
interpreted as the IIB 5-Brane. Like in 10 dimensions we can relate
all these solution directly by T-duality.

\medskip

So far, we have discussed the six--dimensional
$D$-Branes as compactifications of two intersecting 
$D$-Branes in 10 dimensions. The $O(4,20)$ structure was determined
by the structure of the $K3$. To determine the scalar field matrix
${\cal M}$ we have to look for
the heterotic analogue.  Starting from the Type IIA solutions we will
find two heterotic solutions, a pure magnetic (by mapping the
$D$-2-Brane) and one pure electric (by mapping the $D$-0-Brane). By using
the string/string duality transformation we find for the pure
magnetic solution the compactified (magnetic) chiral null model
\cite{be/ka}
\begin{equation}    \label{CNM}
\begin{array}{c}
ds^2_{H}=  ds_3^2 - e^{-4 U} dx^i dx^i 
\quad , \quad  e^{-4U} = e^{4\phi}  \\
 {\cal M}= {\bf 1} + 2 e^{4U} \left( \begin{array} {cc}  \chi^L_{\alpha}
\chi^L_{\beta}&  \chi^L_{\alpha} \chi^R_{\beta}\\
  \chi^R_{\alpha} \chi^L_{\beta} &  \xi \chi^R_{\alpha} \chi^R_{\beta}
  \end{array} \right) \quad ,  \quad
  \left(\begin{array}{c} \vec{F}^{L}_{ij}\\ \vec{F}^{R}_{ij} 
  \end{array} \right) =
  \sqrt{2} \epsilon_{ijm}\partial_m  \left(\begin{array}{c} \vec{\chi}^L \\
  \vec{\chi}^R \end{array}\right)\end{array} \ ,
 \end{equation} 
where $\xi={ |\vec{\chi}^L|^2\over |\vec{\chi}^R|^2}$ and $i = 1,2,3$. 
Uplifted
to 10 dimensions (via $T_4$) one finds
\begin{equation}
\label{d=10magn}
\begin{array}{c}
ds^2_H =  ds_3^2 - e^{-4U} dx^i dx^i 
(dx^{\alpha} + A^{(1)\alpha}_i dx^i)
 \, G_{\alpha\beta}\,(dx^{\beta} + A^{(1)\beta}_i dx^i) \ , \\
 G_{\alpha\beta} = \delta_{\alpha\beta} - \frac{ \chi^L_{(\alpha}
   \chi^R_{\beta)}}{|\vec{\chi}^R|^2 + (\vec{\chi}^R \vec{\chi}^L)}
\quad , \quad  e^{-2 \phi} = e^{2U} \, \frac{1}{\sqrt{\det G}} \ ,
    \end{array}
   \label{monopoles} 
\end{equation}
where $x^{\alpha}$ are the 4 isometry directions. 
The 10d antisymmetric tensor components are given by 
\begin{equation}
B_{\alpha\beta} = \frac{- 2 \chi^L_{[\alpha}
  \chi^R_{\beta]}}{|\vec{\chi}^R|^2 + (\vec{\chi}^R \vec{\chi}^L)}
  \quad , \quad
  B_{\alpha\mu} = A^{(2)}_{\mu\alpha} + B_{\alpha\beta}
  A^{(1)\beta}_{\mu} \ ,
\end{equation}
where $A^{(1/2)}$ are the potentials related to the
original field strength by (without the additional left moving part)
\begin{equation}
\left( \begin{array}{c} \vec{F}^{(1)} \\ -\vec{F}^{(2)} \end{array} \right) =
\frac{1}{\sqrt{2}} \left( \begin{array}{cc} 1 & 1 \\ -1 & 1 \end{array}
\right)
\left(\begin{array}{c} \vec{F^R} \\ \vec{F}^L \end{array} \right)\, .
\end{equation}
The corresponding result for the pure electric case coincides with the
solution in \cite{pe} which is an uplifted version of the solution found in
\cite{se}. This relation to known magnetic and electric solutions
proofs that our $D$-Brane solutions are solutions of the Type
IIA/B effective actions.

On the heterotic side we can interprete the solution as
intersections of known basic $p$-Branes. To make this more transparent
we take only 2 charges and go into the ``(1)/(2)'' basis.
Then the 10d metric and dilaton (\ref{d=10magn}) can be written as
a magnetic chiral null model
 \begin{equation}
ds^2_H = ds_6^2 - H_s \tilde{H}_s dx^i dx^i -
 \frac{H_s}{\tilde{H}_s} \left(dz + {A}^{(1)}_i  d x^i \right)^2
 \quad , \quad    e^{2 \phi} = H_s\, .
\end{equation}
For $\tilde{H}_s=1$ and an appropriate torsion we get the (solitonic)
5-Brane and in the other case ($H_s = 1$) its $T$-dual,
a Taub-NUT soliton. The 24 different charges in 6 dimensions are
related to the different possibilities to choose the isometry
direction $z$ and to the possibility to give the 10-dimensional
solution additional charges with respect to the left moving
sector.

Similarly, if we transform the electric 0-Brane to the heterotic
side we find a solution which is in 10 dimensions an intersection
of a fundamental string (with the harmonic $H_f$) and its 
T-dual, the gravitational wave background (with the harmonic $\tilde{H}_f$)
\begin{equation}
ds^2 = \frac{1}{H_f} (dv du - \tilde{H}_f du^2) - ds^2_8 \quad ,
\quad e^{-2\phi} = H_f \ ,
\end{equation}
which is the electric chiral null model of \cite{Ho1}.

\section{Comments}

We have shown that the elementary $D$--Brane solutions
in six dimensions can be interpreted as the intersection of two 
$D$--Branes in ten dimensions. Furthermore we have shown that
the corresponding heterotic solutions can be viewed as the
intersection of a string (or 5-Brane) with its $T$--dual. It is natural
to extend this analysis and to consider the intersection of more
than two solutions in ten dimensions which could involve both
$D$--Branes and NS/NS solutions. Of special interest are those
solutions which have vanishing scalars upon identification of
the different harmonic functions involved.  

As an example we consider the five--dimensional Reissner-Nordstrom
black hole solution which was considered in \cite{Str1}
to give a microscopic derivation of the entropy in terms of
counting $D$--Brane states. This solution has the metric
\begin{equation}
ds^2 = H^{-2} dt^2 - H ds^2_4 \, , \label{5dRN}
\end{equation}
where $H$ is a harmonic function. A ten--dimensional origin of this
solution has been discussed in \cite{Ca1}. Instead, here we will
discuss its interpretation in terms of six--dimensional solutions.
Given the powers of the harmonic function, it is clear that we should
consider the intersection of a six--dimensional $D$--Brane with a
NS/NS string or its $T$--dual. Requiring that we want to end up with
a constant dilaton in five dimensions restricts ourselves to consider either
the intersection of a 0--Brane with a fundamental string (cancelling
dilatons) or a 1--Brane
with the $T$--dual of a fundamental string (with each vanishing dilaton). 
Thus, in the first case the 6--dimensional intersecting
solution is $1_f \times 0_D$ and is given by
\begin{equation}
ds^2 = \frac{1}{H_f H_0} dt^2 - \frac{H_0}{H_f} ds_1^2 -
  H_0 ds_5^2  \quad , \quad e^{-2 \phi} = \frac{H_f}{H_0} \ ,
\end{equation}
whereas in the second case we have the intersecting solution
$\tilde{1}_f \times 1_D$
\begin{equation}
ds^2 = \frac{1}{H_1} (du dv - \tilde{H}_f du^2) -
      \tilde{H}_1 ds_5^2 \quad , \quad e^{-2 \phi} \sim 1 \ .
\end{equation}
Identifying the harmonic functions we obtain after compactification
over $ds_1$ or $u$, respectively, the metric (\ref{5dRN})
with a vanishing dilaton and constant compactification radii.
Looking on the 10--dimensional origin we find that this
solution is given by 3 intersecting Branes, Type IIA for the first
case and Type IIB for the second case. The Type IIB intersection
has been discussed in \cite{Ca1}.

As a second example for a solution that is interesting in the context
of entropy calculations we discuss the 4--dimensional
Reissner-Nordstrom solution (or their generalizations to more than one
harmonic function).  Again, for this solution their exists a limit
obtained by identification of harmonic functions for which all scalars
dissappear. It is natural therefore to consider it as an elementary
4--dimensional $D$--Brane.  The metric is given by
\begin{equation}
ds^2 = H^{-2} dt^2 - H^{2} ds^2_3 \, . \label{4dRN} 
\end{equation}
It has a natural interpretation in terms of
intersecting $D$--Branes in 6 or 10 dimensions. Since the powers
of the harmonic function in front of the world volume (time) and
the transversal space is the same we can express this solution
directly by $D$--Branes, we do not need the NS/NS--Branes.
As in the case before we have two possibilities. The first
one with cancelling dilaton contributions is given by the
intersection of a $0$-- and 2--Brane in 6 dimensions
\begin{equation}
ds^2 = \frac{1}{H_0 H_2} dt^2 - \frac{H_0}{H_2} ds^2_2 - H_0 H_2
ds^2_3  \qquad , \qquad e^{-2 \phi} = \frac{H_2}{H_0} \ ,
\end{equation}
while the second possibility with each vanishing dilaton is
the intersection of two Dirichlet 1--Branes
\begin{equation}
ds^2 = \frac{1}{H_1 \tilde{H}_1} dt^2 - \frac{H_1}{\tilde{H}_1} ds^2_1 -
\frac{\tilde{H}_1}{H_1} ds^2_1 - H_1 \tilde{H}_1 ds^2_3 
\quad , \quad e^{-2 \phi} \sim 1 \ .
\end{equation}
In analogy with the case of the 5--dimensional Reissner--Nordstrom
black hole we find that upon identification of the harmonic
functions ($H_0$ with $H_2$ or $H_1$ with $\tilde{H}_1$, respectively)
the compactification over the relative transversal space does
not yield additional scalars and is given by (\ref{4dRN}).
Since every 6--dimensional $D$--Brane has an interpretation
as an intersection of two 10--dimensional $D$--Branes we obtain
for this case in 10 dimensions an intersection of 4 $D$--Branes.
Some examples of such an interpretation have been given in
\cite{Ts2}. 

In order to find a non-vanishing area of the horizon (non-vanishing
Bekenstein--Hawking entropy) it is not necessary to identify
the harmonics. The crucial property is, that all scalar fields
stay finite on the horizon ($r=0$). Allowing for any Brane an
independent harmonic function which means an independent charge
yields the general case.

Our classification in terms of intersecting $D$--Branes 
has a natural interpretation in terms of the black hole solutions with
dilaton coupling parameter $a = \sqrt{4/n -1}$ \cite{Pa1}
\footnote{An interpretation of these solutions in terms
of intersecting $M$--Branes has been discussed in \cite{Pa1, Ts2}.}:
\renewcommand{\arraystretch}{1}
\begin{itemize}
\item[1)] $a=\sqrt{3}$ : 10d elementary $D$--Brane;
\item[2)] $a=1$ : 6d elementary $D$--Brane (or 2 intersecting 
 10d $D$--Branes);
\item[3)] $a=1/ \sqrt{3}$: intersection of a 6d $D$--Brane with
a NS/NS Brane (or 3 intersecting 10d Branes);
\item[4)] $a=0$: 4d elementary $D$--Brane (or 2 intersecting 6d $D$--Branes
or 4 intersecting 10d $D$--Branes).
\end{itemize}
So far we have only considered $D$--Branes. It is natural to also
include anti-$D$--Branes which carry the opposite charge.  In the case
that both charges differ only in sign one obtains massless black holes
\cite{Beh1}. This is consistent with the picture that the massless case
in 4 dimensions corresponds to a composition of two or four $a=\sqrt{3}$ 
black holes \cite{Or1}.

\medskip

Finally, we comment on the number of independent harmonic function in
our intersecting solutions. Our basic set of $D$--Branes in 6
dimensions is given by 2 intersecting $D$--Branes in 10 dimensions and
thus contains 2 harmonic functions. The 24 different charges are
related to different radii in the $K3$.  On the heterotic side
(\ref{CNM}) the solution can be naturally extended to more independent
harmonic functions.  At least for the left moving sector we can assume
that the harmonic functions are completely independent yielding 1+4
functions in 6 dimensions (neglecting the 16 additional left moving
modes). On the electric side this case corresponds in 10 dimensions
to the chiral null model that contains more than two independent 
harmonic functions (the $\omega_n$ functions in \cite{Beh1}).
These additional harmonic functions are related to momentum modes
in the internal directions. In analogy, the magnetic chiral null
model can be described by the same number of independent harmonic
functions. It would be interesting to
see what the role of the additional harmonic functions are in the
$D$--Brane picture.

\vspace{.5truecm}

\noindent Note added: Soon after the appearence of this paper there
appeared a preprint \cite{GKT} which has some overlap with the present work.

\vspace{.5truecm}

\noindent {\bf Acknowledgements}
\vspace{.5truecm}

We thank the referee for urging us to consider the supersymmetry of the
intersecting $D$--Brane configurations and Arkady Tseytlin for pointing
out to us that the $0||2$ and $1||3$ configurations of the $n=1$ class
are not solutions. E.B.~thanks Mees de Roo for useful discussions.
The work of K.B.~is supported by the DFG.  He would like to thank
the Institute for Theoretical Physics of Groningen University for its
hospitality.  The work of E.B.~has been made possible by a fellowship
of the Royal Netherlands Academy of Arts and Sciences (KNAW).  He
thanks the Institute for Theoretical Physics of
Humboldt University Berlin for its hospitality.  The work of B.J.~was
performed as part of the research programme of the ``Stichting voor
Fundamenteel Onderzoek der materie'' (FOM).

\vfill\eject

\thispagestyle{empty}
\textwidth 170mm
\textheight 220mm
\topmargin -15mm
\oddsidemargin 0mm
\evensidemargin 0mm


\begin{figure}[t]
\begin{center}
\mbox{\epsfig{file=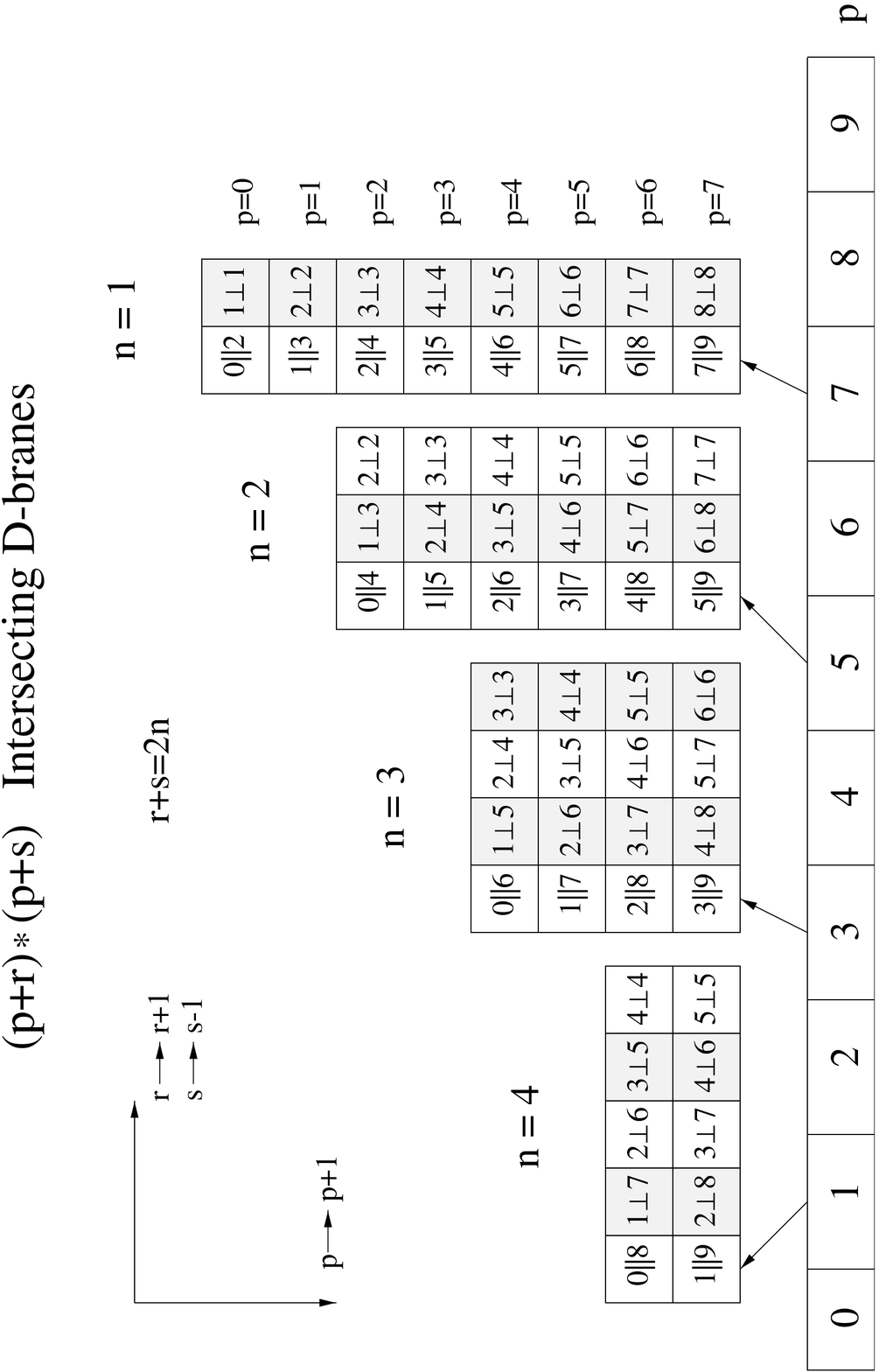, width=130mm}}
\end{center}
\noindent {\bf Table}\ \ {\it The table contains all possible intersecting
configurations of two $D$--Branes in ten dimensions. 
The notation is explained in
section 2. 
Note that each class starts at the top with $p=0$ intersection
and that $p$ is constant inside the horizontal line of a given class.
}
\end{figure}

\end{document}